\documentclass[a4paper,10pt]{article}
\usepackage[utf8]{inputenc}
\usepackage[hidelinks]{hyperref}
\usepackage{csquotes}
\usepackage[english]{babel}
\usepackage{authblk}
\usepackage[font=small,labelfont=bf]{caption}
\usepackage{algorithm}
\usepackage{amssymb}
\usepackage{algpseudocode}
\usepackage{amsmath}
\usepackage{placeins}
\usepackage{graphicx}
\usepackage{titlesec}
\usepackage{changepage}
\floatstyle{plain}
\newfloat{fig}{htbp}{lop}
\floatname{fig}{Fig.}
\graphicspath{ {algo/} }
\date{}
\newcommand{\ts}{\textsuperscript}

\font\myfont=cmr12 at 14pt
\titleformat{\section}{\bfseries\fontsize{12pt}{17pt}\selectfont}{\thesection}{1em}{}
\titleformat{\subsection}{\bfseries\fontsize{10pt}{14pt}\selectfont}{\thesubsection}{1em}{}
\titleformat{\subsubsection}[runin]{\bfseries\fontsize{10pt}{12pt}\selectfont}{\thesubsubsection}{1em}{}

\makeatletter
\newenvironment{breakablealgorithm}
  {
   \begin{center}
     \refstepcounter{algorithm}
     \hrule height.8pt depth0pt \kern2pt
     \renewcommand{\caption}[2][\relax]{
       {\raggedright\textbf{\fname@algorithm~\thealgorithm} ##2\par}%
       \ifx\relax##1\relax 
         \addcontentsline{loa}{algorithm}{\protect\numberline{\thealgorithm}##2}%
       \else 
         \addcontentsline{loa}{algorithm}{\protect\numberline{\thealgorithm}##1}%
       \fi
       \kern2pt\hrule\kern2pt
     }
  }{
     \kern2pt\hrule\relax
   \end{center}
  }
\makeatother
\DeclareMathOperator{\arctantwo}{arctan2}
\usepackage{color}
\usepackage{listings}
\definecolor{GrayCodeBlock}{RGB}{241,241,241}
\definecolor{BlackText}{RGB}{110,107,94}
\definecolor{AquaTypename}{RGB}{45, 191, 184}
\definecolor{BrownString}{RGB}{195,105,57}
\definecolor{AquaKeyword}{RGB}{45, 191, 184}
\definecolor{BlueKeyword}{RGB}{54, 124, 208}
\definecolor{PurpleKeyword}{RGB}{184,84,212}
\definecolor{GrayComment}{RGB}{170,170,170}
\definecolor{GoldDocumentation}{RGB}{180,165,45}
\lstdefinelanguage{rust}
{
    columns=fullflexible,
    keepspaces=true,
    frame=single,
    framesep=0pt,
    framerule=0pt,
    framexleftmargin=4pt,
    framexrightmargin=4pt,
    framextopmargin=5pt,
    framexbottommargin=3pt,
    xleftmargin=4pt,
    xrightmargin=4pt,
    backgroundcolor=\color{GrayCodeBlock},
    basicstyle=\ttfamily\color{BlackText},
    keywords={
        true,false,
        unsafe,async,await,move,
        use,super,self,mod,
        enum,fn,const,static,let,mut,ref,type,impl,dyn,trait,where,as,
        break,continue,if,else,while,for,loop,match,return,yield,in
    },
    keywordstyle=\color{AquaKeyword},
    keywordstyle=[2]{
        pub,struct,extern,crate
    },
    keywordstyle=[2]\color{BlueKeyword},
    ndkeywords={
        bool,u8,u16,u32,u64,u128,i8,i16,i32,i64,i128,char,str,
        Self,Option,Some,None,Result,Ok,Err,String,Box,Vec,Rc,Arc,Cell,RefCell,HashMap,BTreeMap,
        FeedMessage,UserFeed,
        macro_rules
    },
    ndkeywordstyle=\color{AquaTypename},
    comment=[l][\color{GrayComment}\slshape]{//},
    morecomment=[s][\color{GrayComment}\slshape]{/*}{*/},
    morecomment=[l][\color{GoldDocumentation}\slshape]{///},
    morecomment=[s][\color{GoldDocumentation}\slshape]{/*!}{*/},
    morecomment=[l][\color{GoldDocumentation}\slshape]{//!},
    morecomment=[s][\color{AquaTypename}]{\#![}{]},
    morecomment=[s][\color{AquaTypename}]{\#[}{]},
    stringstyle=\color{BrownString},
    string=[b]"
}

\title{\myfont {\textbf{Algorithmic Analysis of GTFS-RT vehicle position accuracy}}}
\author[1,2]{
\small Joshua Wong
}
\affil[1]{\footnotesize California State Polytechnic University, Humboldt, Arcata CA 95521, USA}
\affil[2]{\footnotesize Lavender Computing Collective, West Covina, CA 91770, USA}

\affil[ ]{\textit {jw504@humboldt.edu}}
\begin{document}

\maketitle
\begin{adjustwidth}{1cm}{1cm}
        \textbf{Abstract.}
This paper presents three novel algorithms for calculating geodesic intersections on an ellipsoid. These algorithms are applied in a case study analyzing real-time transit data in California to assess vehicle position drift. The analysis reveals that while certain data anomalies can be corrected, large-scale discrepancies persist. The paper concludes by proposing a set of practical solutions that can be implemented by either data producers or consumers to significantly improve positional accuracy.\end{adjustwidth}
\vspace{1pt}
\begin{adjustwidth}{1cm}{1cm}
\end{adjustwidth}

\section{Introduction}
Many transit providers are adopting the General Transit Feed Specification (GTFS) in order to provide machine-readable data to transit application providers. GTFS can include data such as routes, schedules, and GTFS also has a real-time component, named GTFS-RT (real-time). This specification contains time-sensitive data such as vehicle position and vehicle occupancy. However, GTFS and GTFS-RT data have little research done on their accuracy. Using this data, we can determine the accuracy of the real-time data. In this paper, I will measure the vehicle position accuracy of California Transit Agencies, as it relates to their scheduled routes.

\section{GTFS and GTFS-RT}
The General Transit Feed Specification has two components, the schedule and the real-time. The GTFS schedule component consists of text files in a CSV format that each store data about a particular component of a transit agency's schedule \cite{gtfs_schedule_2024}. The text files used in this study, inside a GTFS feed, are:
\begin{enumerate}
\item shapes.txt: The shapes of transit routes are formatted as linestrings. \\Linestrings are ordered sets of points over which a vehicle travels in order. These shapes are related to trips by the trip ID.
\item trips.txt: All trips for an agency. Each trip includes data about the trip's ID, route ID, service ID, name, shape ID, along with some other fields. This file is a lookup for relationships between many data entries via ID. For example, the shape of a trip can be found just by looking at this file and searching the corresponding shape ID in the shapes.txt file.
\item routes.txt: All routes an agency operates. Each route contains data such as name, type, and branding, along with some other fields.
\item agency.txt: Contains basic information about the agencies that manage the GTFS feed, such as name, website, timezone, language, and contact information.
\end{enumerate}
The rest of the GTFS feed file descriptions can be found at \\ \href{https://gtfs.org/schedule/reference}{gtfs.org/schedule/reference}.

The second component of the General Transit Feed Specification is the real-time feed specification \cite{gtfs_realtime_2024}. This feed consists of three separate data entities. They are:
\begin{enumerate}
\item Trip Updates: Data about the real-time arrivals and departures of a trip, along with the progress of a trip. This feed sometimes contains the This feed can be updated with new trips or removed trips if necessary.
\item Vehicle Positions: Data about the position of a vehicle. Can contain identifying information such as vehicle ID and trip ID.
\item Alerts: Data about an incident in the public transit network.
\end{enumerate}
This study uses Vehicle Positions data entities, as they contain the WGS84 coordinates used to locate the vehicle.

The World Geodetic System (WGS84) is the standard used when it comes to
cartography and navigation, as well as geodesy. WGS84 uses the GRS80 ellipsoid as the
basis for its parameters \cite{fissgus2013difference}. We will use the WGS84 coordinate system and represent the Earth as an oblate spheroid. The difficulty in this task comes from the issue of the Earth's irregular shape.

Finally, I use OpenStreetMap (OSM) data to map out the results of this study, as OpenStreetMap provides a free data source for geographic features, such as roads, bus stops, and train stations \cite{biljecki2023quality}.
Using all of these data formats, I conduct a study using California Transit Agency Data.
%
\section{Literature Review}
\subsection{GTFS and GTFS-RT performance}
The most studied metric in GTFS-RT data is transit delay performance. Aemmer et al. \cite{aemmer2022measurement}, Lopes \cite{lopespondionstracker}, and Steiner et al. \cite{steiner2015quality}, measure transit performance using GTFS-RT, using the metric of transit delay compared to GTFS schedule data. Aemmer uses the one year of GTFS-RT data scraped from the King County Metro bus network in Seattle,
Washington, while Stenier uses the Netherlands. Stenier discovers that the feeds in the Netherlands are missing a large majority of vehicle position and trip delay data. In fact, 43\% of VehiclePositions data packets near a station did not have a delay value.

Wessel's thesis \cite{wessel_2019} used speed to detect positional real-time data errors between location reports. However, this method does not measure the amount of GPS positional error. To correct errors like this, along with the general amount of GPS noise, Wessel used map-matching against OpenStreetMap data using Open Source Routing Machine (OSRM). OSRM uses a probabilistic map-matching algorithm, returning the resulting route along with the probability that it is correct using a naive Bayes classifier.

Another important metric is the quality of the GTFS data and what missing fields it may contain. Barbeau and others \cite{barbeau2018overcoming} conducted a study using data from regional transit in the Tampa Bay area in Florida. The authors state that both the GTFS schedule and GTFS-RT having missing data fields may have a negative effect on an agency's ridership, along with the riders' opinion on an agency, as they have less data in order to go and use public transit. Furthermore, one of the problems they faced was bad shape data in GTFS schedule data, such that the actual travel path of the bus was significantly different from the shape recorded in the shapes.txt file. This led to Barbeau et. al. to create an application named GO\_Sync to correct agency data, such as bus stops. However, automating synchronization of way/shape data is not practical due to the change of the characteristics of a street \cite{barbeau2011enabling} \cite{tran2013go_sync}. However, some parts of synchronization between GTFS shape data and OSM data can be automated, such as using ArcGIS for merging.

Devunuri and Lehe \cite{devunuri2024survey} conducted a ``A Survey of Errors in GTFS Static Feeds from the United States''. They concluded that the two most common errors as a result of an incorrect shapes.txt were:
\begin{enumerate}
\item \textbf{equal\_shape\_distance\_diff\_coordinates:} Two points on a route shape have the same shape\_dist\_traveled but different coordinates (which is impossible).
\item \textbf{trip\_distance\_exceeds\_shape\_distance:} The maximum of shape\_dist\\\_traveled in stop\_times.txt exceeds the maximum of shape\_dist\_traveled in shapes.txt.
\end{enumerate}
In particular, these errors have to do with the optional field of shape\_dist\_traveled (51\%) of all error occurrences.
\subsection{Geodesic Intersections and Methods}
There are two major problems in geodesy, the direct and inverse problem \cite{rainsford1955long}.
The direct problem states, given the geographical coordinates of one point and the distance and azimuth from it to another, find the coordinates of the second point. 
The inverse problem is the reverse of that, stating, given the geographical coordinates of two points, find the distance and azimuths between them.
Karney's solutions \cite{Karney_2012} for the direct and inverse geodesy problem are the most accurate, and are better than the previous work \cite{vincenty1975direct} in three ways.
\begin{enumerate}
	\item The accuracy is increased to match the standard precision of most computers.
	\item Karney's solution converges for all pairs, compared to Vincenty’s method, which fails to converge for nearly antipodal points.
	\item Differential and integral properties of the
	geodesics are computed.
\end{enumerate}
Therefore, they are the basis for many geodesic algorithms, including my algorithms later in this paper.

Baselga and Jos\'e Carlos Mart\'inez-Llario's (BML) \cite{Baselga2018} most relevant method is for minimum point-to-line distance on an ellipsoid. They use ``direct and inverse problems of geodesy plus the necessary condition of intersection at right angles``, resulting in the algorithm below.
\begin{breakablealgorithm}
	\caption{Baselga and Jos\'e Carlos Mart\'inez-Llario's minimum point-to-line distance on an ellipsoid algorithm}
	\label{alg:BML}
	\begin{algorithmic}[1]
		\Require $A, B, C$ with coordinates $(lat_A, lon_A), (lat_B, lon_B), (lat_C, lon_C)$
		\Ensure Intersection Point, Distance, Azimuth $(lat, long, d, \alpha)$
		\State $R \gets 6378137$ \Comment{radius (in m) of the WGS84 ellipsoid at the equator}
		\State $s_{ax} \gets \infty$
		\Loop \\
		\Comment{the inverse function from \cite{Karney_2012} also returns the values: distance, azimuth, reduced length of geodesic, geodesic scale}
		\State $(s_{ac}, \alpha_{ac}, m_{ac}, M_{ac})\gets$ inverse$(lat_A, lon_A, lat_C, lon_C)$
		\State $\alpha_{ab} \gets$ inverse$(lat_A, lon_A, lat_B, lon_B)$
		\State $\alpha = \alpha_{ac} - \alpha_{ab}$
		\State $s_{px} \gets R * \arcsin(\sin(s_{ap}/R)*\sin{\alpha})$
		\State $s_{ax} \gets 2 * R * \arctan\left(\dfrac{\sin{\frac{90+\alpha}{2}}}{\sin{\frac{90-\alpha}{2}}}*\tan{\dfrac{s_{ap} - s_{px}}{2R}}\right)$
		\Comment{The direct method in \cite{Karney_2012} returns a point given distance and azimuth}
		\State $(lat_n, lon_n) \gets$ direct$(lat_a, long_a, \alpha_{ab}, s_{ax})$ \\
		\Comment{Adjust $10^{-2}$ to either increase accuracy or computation speed}
		\If{$\lvert s_{ax} \rvert < 10^{-2}$}
		\Return $(lat_a, lon_a, s_{ac}, \alpha_{ab})$
		\Else
		\State $A \gets (lat_n, lon_n)$
		\EndIf
		\EndLoop
	\end{algorithmic}
\end{breakablealgorithm}
Currently, Karney's geodesic intersection algorithms \cite{karney2023geodesic} are the most accurate methods. They include the intersection of geodesics, geodesic segments, the next closest intersection, and all geodesic intersections. However, the most relevant method is his optimization of BML's \cite{Baselga2018} method for minimum point-to-line distance on an ellipsoid, where Karney gives two improvements to their algorithm. Karney replaces the iterative part of their algorithm with these equations.
\begin{align}
	s_{ax} &= R * \arctantwo((\sin(s_{ac} / R) * \cos a), \cos(s_{ac} / R)) \\
	s_{ax} &= \dfrac{m_{ap} * \cos a}{(m_{ac} / s_{ac}) \cos^2{a} + M_{ac} \sin^2{a}}
\end{align}
These equations lower the convergence of the BML algorithm \ref{alg:BML} to less than quadratic. In reality, due to the extremely quick convergence of this method, there will only be several cases where there will be a significant speedup.
\section{Methods}
\subsection{Point-Geodesic Segment Problem}
The problem states: Given a geodesic segment, defined by the shortest path between two points in a surface, and an arbitrary point on the surface, find the closest point on that line to that arbitrary point, along with the minimum distance.
This problem is used to calculate the exact position along the route where a vehicle is located between two points within the route's shape.
\subsubsection*{Solution on an ellipsoid}
I used the algorithm from Baselga and Martinez-Llario
\cite{Baselga2018}, along with improvements from Karney \cite{karney2023geodesic}. However, BML's algorithm fails if the point cannot create an intersection at right angles with the geodesic. Therefore, I used Karney's methods in order to calculate the azimuths \cite{Karney_2012}, and then I modulo them in order to get the internal angles of the ellipsoidal triangle.
My contribution to this algorithm solves for all arbitrary points directly on the geodesic segment.
\begin{breakablealgorithm}
\caption{Calculation of an ellipsoidal triangle's internal angles}\label{alg:angle}
\begin{algorithmic}[1]
\Require $A, B, C$ with coordinates $(lat_A, lon_A), (lat_B, lon_B), (lat_C, lon_C)$
\Ensure Angles: $(\alpha, \beta, \gamma)$ \\
\Comment{inverse function is from \cite{Karney_2012} and returns the forward-facing azimuths}
\State $\alpha_{ab} \gets $inverse$(lat_A, lon_A, lat_B, lon_B)$
\State $\alpha_{bc} \gets $inverse$(lat_B, lon_B, lat_C, lon_C)$
\State $\alpha_{ca} \gets $inverse$(lat_C, lon_C, lat_A, lon_A)$
\State $\alpha \gets (\alpha_{ca} - \alpha_{ab}) \pmod{360}$
\State $\beta \gets (\alpha_{ab} - \alpha_{bc}) \pmod{360}$
\State $\gamma \gets (\alpha_{bc} - \alpha_{ca}) \pmod{360}$ \\
\Return $(\alpha, \beta, \gamma)$
\end{algorithmic}
\end{breakablealgorithm}
This geodesic intersection algorithm does not calculate the solution if the point given to intersect the geodesic cannot create a perpendicular angle with the geodesic segment. The checking for that is done at the point-linestring problem level, as comparing the results of this algorithm is much more efficient there. Furthermore, the first check in the algorithm currently only works for small ellipsoidal triangles as it assumes the maximum sum of internal angles of the ellipsoidal triangle is 180$^{\circ}$. This worked for my analysis as the majority of data points were relatively close to each other. As a consequence of this property, many of the calculations converged with one iteration with an accuracy of $10^{-2}$ meters.
\begin{breakablealgorithm}
\caption{Calculation of a geodesic intersection}
\label{alg:intersection}
\begin{algorithmic}[1]
\Require $A, B, C$ with coordinates $(lat_A, lon_A), (lat_B, lon_B), (lat_C, lon_C)$
\Ensure Intersection Point, Distance, Azimuth $(lat, long, d, \alpha)$
\State $R \gets 6378137$ \Comment{radius (in m) of the WGS84 ellipsoid at the equator}
\State angles $\gets$ alg\ref{alg:angle}$(A, B, C)$
\If{$180 \in$ angles}
	\Return{$(lat_C, lon_C, 0, 0)$}
\Else
	\State iter\_num $\gets 0$
	\State $s_{ax} \gets \infty$
\Loop \\
	\Comment{the inverse function from \cite{Karney_2012} also returns the values: distance, azimuth, reduced length of geodesic, geodesic scale}
	\State $(s_{ac}, \alpha_{ac}, m_{ac}, M_{ac})\gets$ inverse$(lat_A, lon_A, lat_C, lon_C)$
	\State $\alpha_{ab} \gets$ inverse$(lat_A, lon_A, lat_B, lon_B)$
	\State $\alpha = \alpha_{ac} - \alpha_{ab}$
	\If{iter\_num = 0}
		\State $s_{ax} \gets R * \arctantwo((\sin(s_{ac} / R) * \cos \alpha), \cos(s_{ac} / R))$
	\Else
		\State $ s_{ax} \gets \dfrac{m_{ap} * \cos \alpha}{(m_{ac} / s_{ac}) \cos^2{\alpha} + M_{ac} \sin^2{\alpha}}$
	\EndIf \\
	\Comment{The direct method in \cite{Karney_2012} returns a point given distance and azimuth}
	\State $(lat_n, lon_n) \gets$ direct$(lat_a, long_a, \alpha_{ab}, s_{ax})$ \\
	\Comment{Adjust $10^{-2}$ to either increase accuracy or computation speed}
	\If{$\lvert s_{ax} \rvert < 10^{-2}$}
		\Return $(lat_a, lon_a, s_{ac}, \alpha_{ab})$
	\Else
		\State $A \gets (lat_n, lon_n)$
		\State iter\_num $\gets$ iter\_num + 1
	\EndIf
\EndLoop

\EndIf
\end{algorithmic}
\end{breakablealgorithm}
\subsection{Point-Linestring Problem}
The problem states: Given a Linestring, defined by a set of ordered points, and a random point, find the closest point in that set to that random point. Prove that this point and one of the neighboring points create a line segment or geodesic that contains a point such that the point is the closest part of the linestring to that random point.
This problem is used to calculate how far and where on the shape of the route the route's vehicle is near.
\subsubsection*{Solution}
My novel solution uses a k-d tree in order to make the linestring's points, $S$, searchable by \textit{k} nearest neighbor. This allows for an $O(\log n)$ computation, compared to an $O(n)$ brute force solution of computing all the minimum point to geodesic problems for all geodesic segments in a linestring. After finding the nearest point in the linestring from the random point, $p$, I used the neighbors of that point in the linestring to create two geodesics, $\overline{S_pS_{p+1}}$ and $\overline{S_pS_{p-1}}$. Apply the BML Algorithm to $p$ and the geodesics. Return the shortest distance between the results and the distance of the nearest neighbor point in the linestring.
\begin{breakablealgorithm}
\caption{Point to linestring intersection algorithm}
\label{alg:ls}
\begin{algorithmic}[1]
\Require $P: (lat_p, lon_p), L:[(lat_n, lon_n), (lat_{n+1}, lon_{n+1}), (lat_{n+2}, lon_{n+2})...]$
\Ensure Intersection Point, Distance $(lat, long, d)$
\State kdtree $\gets$ new(kdtree)
\For{$i \in$ 1..len(L)}
	\State insert(kdtree, $(i , L[i])$)
\EndFor \Comment{In actual code, the k-d tree is often pregenerated, so this step is often skippable}
\State nearest\_neighbor $\gets$ findnn(kdtree, $P$)
\State geodesic$_{1} \gets $ (nearest\_neighbor[point], L[nearest\_neighbor[i-1]])
\State geodesic$_{2} \gets $ (nearest\_neighbor[point], L[nearest\_neighbor[i+1]])
\State dist $\gets$ [(nearest\_neighbor[point], nearest\_neighbor[distance])]
\State insert(dist, BML(geodesic$_1$, P))
\State insert(dist, BML(geodesic$_2$, P)) \\
\Return min(dist[distance])
\end{algorithmic}
\end{breakablealgorithm}
\subsection{Data Gathering}
\label{data}
The data used in the analysis was gathered by the California Integrated Travel Project (Cal-ITP), using the URL sources from \cite{gtfs_datasets}.
Cal-ITP uses a serverless system to request GTFS-RT data from around 750 real-time feeds and 135 transit agencies. Every twenty seconds, a request is sent to all of these real-time feeds, and a GTFS-RT FeedMessage is returned and stored as a protobuf binary. A similar system is used for the GTFS Static data, but data is fetched daily at midnight local time instead of three times a minute.
\section{Analysis and Results}
\label{analysis}
The GTFS-RT dataset consisted of around 75 million FeedMessages with 460 million data points, while the GTFS dataset consisted of approximately 1700 individual agency/agency group GTFS datasets. This data was collected throughout the first week of July, from July 1\ts{st} to July 7\ts{th}, 2024, using the methods described in section \ref{data}. Using a Rustlang implementation of algorithms \ref{alg:angle}, \ref{alg:intersection}, and \ref{alg:ls}, I analyzed the whole dataset and returned approximately 314 million intersections. Another relation required to analyze VehiclePositions was to get their related routes. This was done by getting the trip\_id from the data points vehicle\_id, then using that trip\_id to get the related shape\_id. However, this causes the most errors to occur, as many FeedMessages had the following issues.
\begin{enumerate}
  \item Missing GTFS: ~4,800,000 FeedMessages
  \item No vehicle/vehicle\_id: ~31,000 data points
  \item No position data: ~530,000 data points
  \item No related trip/shape id: ~132,000,000 data points
\end{enumerate}
This raises a concern for further analysis, as around 30\% of the dataset had errors that made doing analysis impossible. However, Steiner et al. \cite{steiner2015quality} wrote that approximately 43\% of their real-time dataset was missing fields required for their analysis.
\subsection{Nightly Pattern}
\begin{figure}[htbp]
    \centering
    \includegraphics[width=\textwidth]{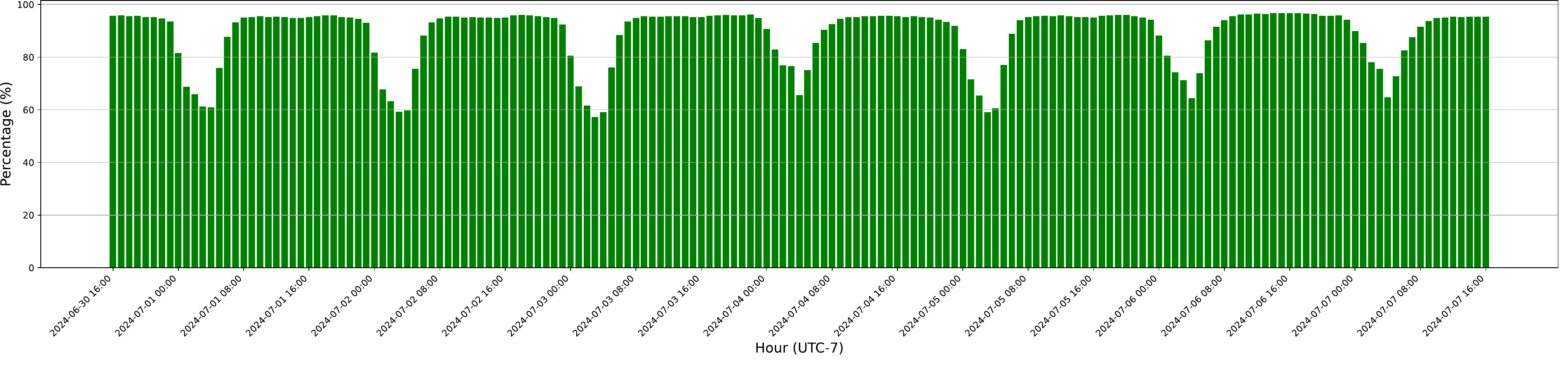}
    \caption{Percentage of Vehicles within 35 meters of their scheduled route.}
    \label{fig:combined_hourly_percent_within_35m}
\end{figure}
\begin{figure}[htbp]
    \centering
    \includegraphics[width=\textwidth]{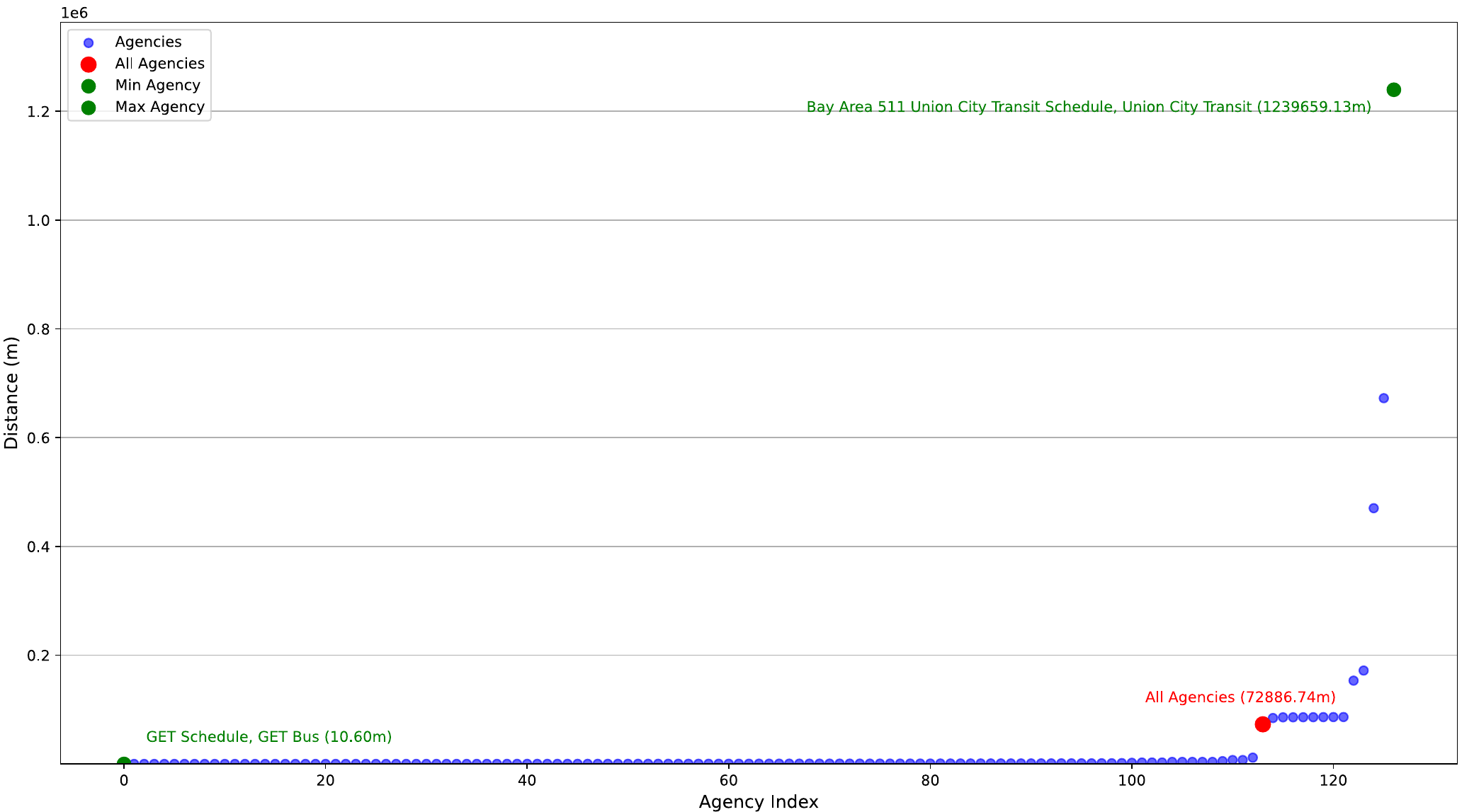}
    \caption{Per agency's standard deviation of vehicle distance from route.}
    \label{fig:standard_deviation}
\end{figure}

\FloatBarrier
Figure \ref{fig:combined_hourly_percent_within_35m} has a nightly pattern of the percentage of vehicles within 35m of their route going down. 35 meters was chosen as it is the size of a fairly large city road, such that any vehicle beyond 35 meters would most likely be an error. This is due to some public transit agencies in California not unlinking their vehicles from their trips while in storage in a faraway lot. Furthermore, these agencies sometimes have vehicles that also do not disable their transponders in maintenance lots.
\begin{figure}[htbp]
    \centering
    \includegraphics[width=\textwidth]{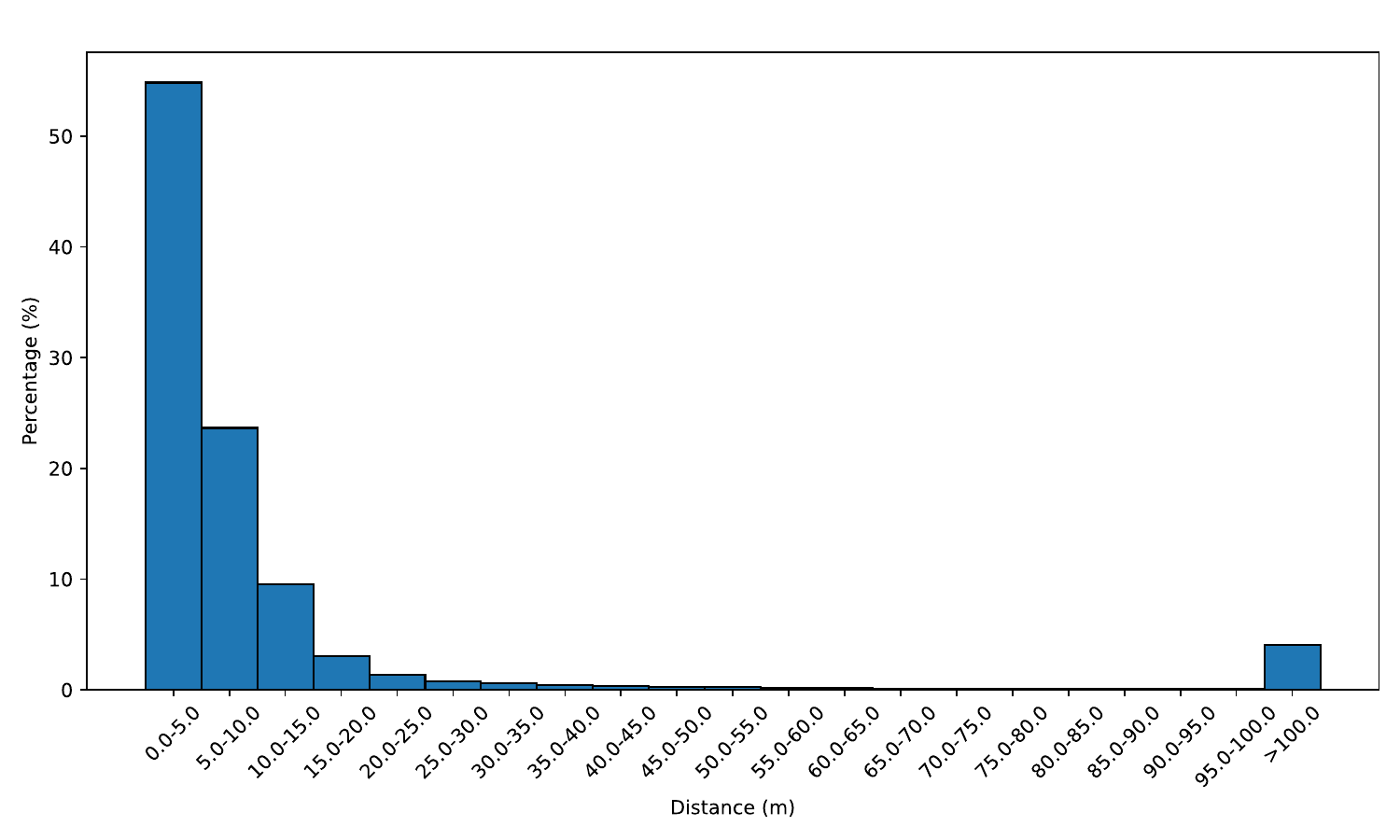}
    \caption{Distribution of vehicle distance from scheduled route}
    \label{fig:distance_percentage_distribution}
\end{figure}
\FloatBarrier
Another issue that may have potentially caused the high standard deviation \ref{fig:standard_deviation} found in the dataset is mentioned by Devunuri et al. \cite{devunuri2024survey}. They mention that a similar error, stop\_too\_far\_from\_shape, is something that is possible throughout GTFS. Therefore, the GTFS dataset I am using as a reference point may have caused the unusually high standard deviation, as in service, a transit vehicle may be running in its correct route, but programmatically, the data has not been updated, causing issues such as a high standard deviation.

\FloatBarrier
\subsection{Map}

\begin{figure}[htbp]
    \centering
    \includegraphics[width=6cm]{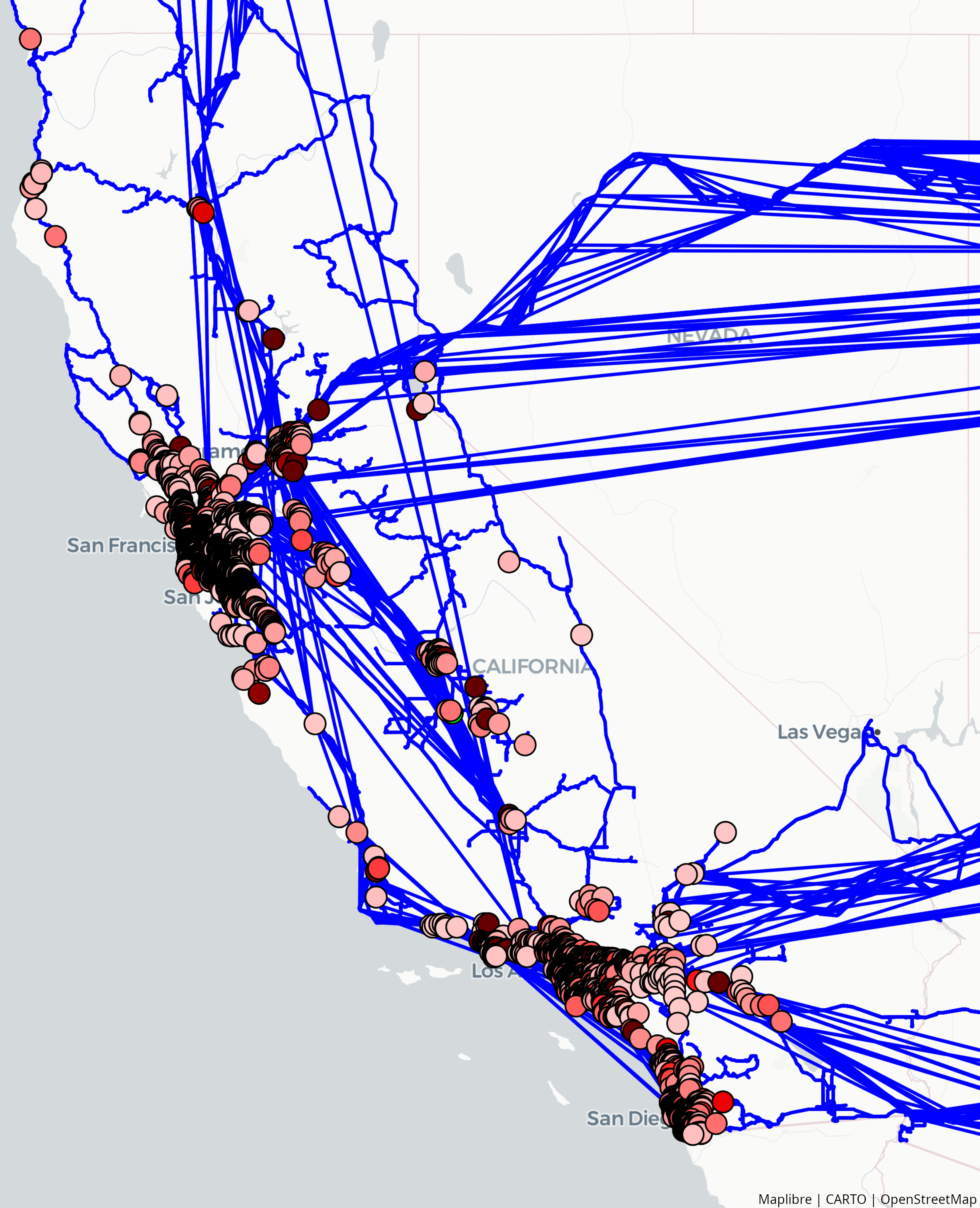}
    \caption{Map of California's GTFS and GTFS-RT}
    \label{fig:cal}
\end{figure}
Figure \ref{fig:cal} is a systemwide map of the dataset analyzed in this paper. Most of the data seems to be coming from the San Francisco Bay Area and Los Angeles County. The GTFS data in relation to OpenStreetMap was relatively accurate, enough to map California. However, some routes in GTFS were 10-30 meters away from a road/pathway.
\begin{figure}[htbp]
    \centering
    \includegraphics[width=12cm]{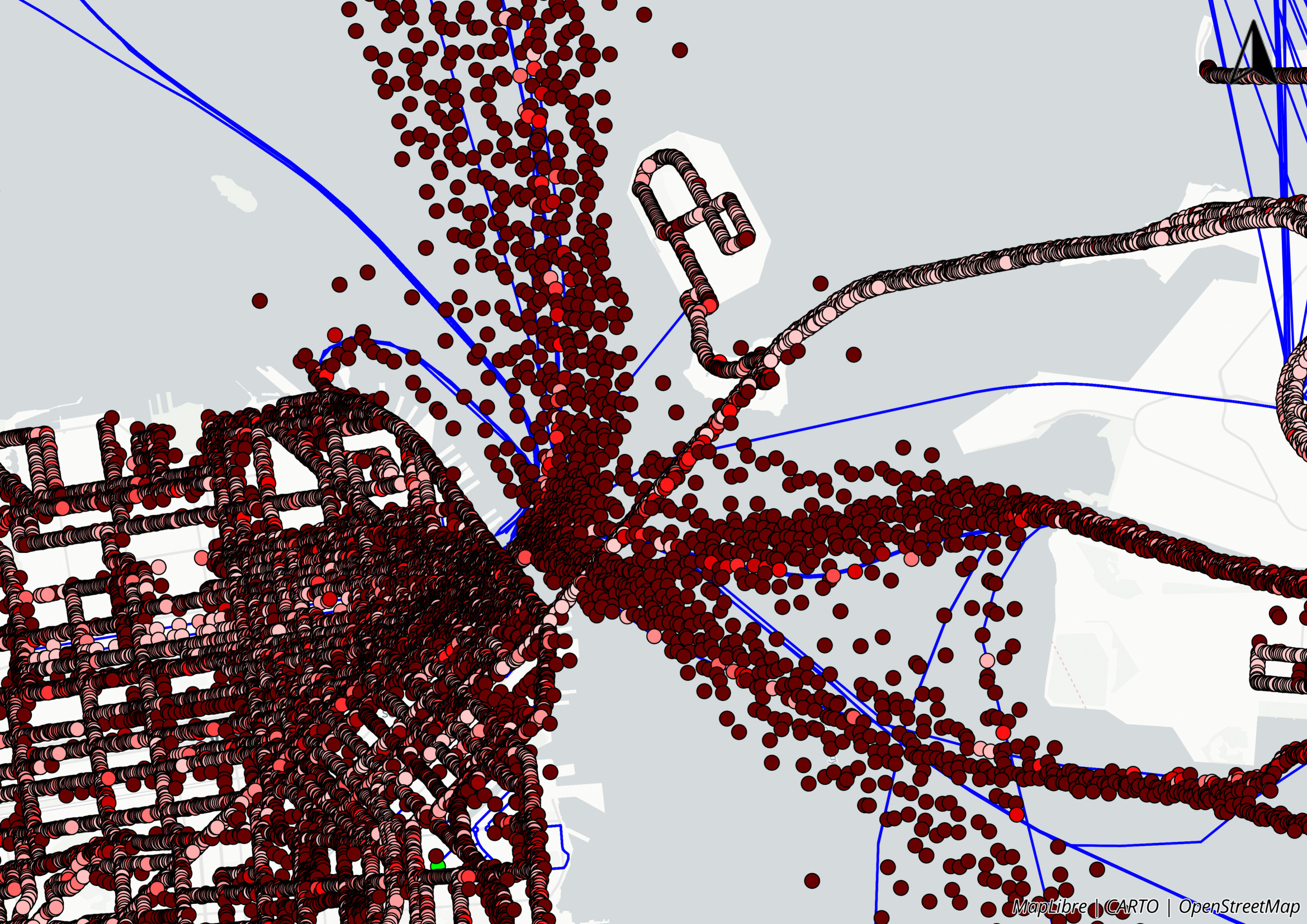}
    \caption{Map of GTFS-RT Errors over the San Francisco Bay water}
    \label{fig:bay}
\end{figure}
\FloatBarrier
Another thing noticed with the map is the large number of VehiclePositions in the water. This could be due to the weakness of signals over the body of water, or it could be due to underground routes that trains take underground.
\section{Discussion}
A Proof for the calculation of internal angles of an ellipsoidal triangle \ref{alg:angle} could potentially be done using Legendre's theorem on spherical triangles \cite{1982224} and a geometric proof such as Ballantine's and Jerbert \cite{2306514}. Furthermore, a faster (linear time) algorithm for the point-linestring problem is possible, as the creation of the k-d tree is $O(n\log n)$. One method used in Georust, a community-run ecosystem of geospatial tools, is a brute-force calculation of all geodesics in the linestring with a distance algorithm. However, the k-d tree was used many times here, which made the exchange of an $O(n\log n)$ preprocessing step for the sake of an $O(\log n)$ distance calculation, compared to the $O(n)$ brute force solution. Finally, the k-d tree method used in this paper is not proven to guarantee the shortest possible distance between a point and a linestring. This must be proven later. An example of a potentially problematic scenario is a linestring that corresponds to a harmonic lateral displacement, such as sinusoidal offset, to a great circle, along with a point with a zero or small distance from the pole of the great circle. This will create many equidistant points on the linestring to the point.

Another issue that could be solved is the number of GTFS-RT errors that came up during analysis. There could be potentially more ways to relate routes to VehiclePositions other than the method mentioned in section \ref{analysis}. Furthermore, methods used in Wessel et al. \cite{WESSEL201792} to dynamically update GTFS in order ``to store the observed rather than the scheduled transit operations'' might help increase. However, this will not help the agencies that do not detach their vehicles from the trip ID once the trip is completed, which seemed to be the cause for the majority of the inaccurate points.
\section{Conclusion}
In this work, I created three novel algorithms in order to calculate the minimum geodesic distance from point to linestring. Starting from correcting BML's algorithm to work for geodesic segments, implementing Karney's improvements to BML's algorithm, and building an algorithm to calculate from point to linestring. From those algorithms, I connected public transit routes to their shapes and analyzed the real-time vehicle position data's accuracy against the static shape data. Finally, I created a map to display the results of my algorithms.
\section{Acknowledgements}
The author would like to thank Kin Tsang of Lavender Computing Collective and a student at California State University, Los Angeles, for providing access to computational resources. I would also like to thank Evan Siroky, Research Data Manager, at the California Department of Transportation, for providing a dataset for my analysis.
\bibliographystyle{plain}        
\bibliography{algo}        
\end{document}